\documentclass[a4paper,fleqn]{cas-sc}
\usepackage[authoryear,longnamesfirst]{natbib}
\usepackage{enumitem}
\usepackage{etoolbox}
\usepackage{amsmath}
\def\tsc#1{\csdef{#1}{\textsc{\lowercase{#1}}\xspace}}
\tsc{WGM}
\tsc{QE}
\tsc{EP}
\tsc{PMS}
\tsc{BEC}
\tsc{DE}
\DeclareMathOperator*{\argmin}{arg\,min}
\newtheorem{theorem}{Theorem}
\newtheorem{lemma}[theorem]{Lemma}
\newdefinition{rmk}{Remark}
\newproof{pf}{Proof}
\newproof{pot}{Proof of Theorem \ref{thm}}
\def\multiset#1#2{\ensuremath{\left(\kern-.3em\left(\genfrac{}{}{0pt}{}{#1}{#2}\right)\kern-.3em\right)}}
\begin{document}

\let\WriteBookmarks\relax
\def\floatpagepagefraction{1}
\def\textpagefraction{.001}

\newlist{algolist}{enumerate}{3}
\setlist*[algolist]{leftmargin=*,before={\apptocmd{\item}{\mbox{}}{}{}}}
\setlist*[algolist,1]{label={\itshape Step \arabic*:}}
\setlist*[algolist,2]{label={\itshape Step \arabic{algolisti}.\arabic*:}}
\setlist*[algolist,3]{label={\itshape Step \arabic{algolisti}.\arabic{algolistii}.\arabic*:}}
\shorttitle{A Novel exact algorithm for economic lot-sizing with piecewise
linear production costs}

\shortauthors{K.Papadopoulos}

\title [mode = title]{A Novel exact algorithm for economic lot-sizing with piecewise
linear production costs}                      



%
\author[1,3]{Kleitos Papadopoulos}[
                        orcid=0000-0001-7511-2910]


\fnmark[1]

\ead{leitospa@gmail.com}


\credit{Conceptualization of this study, Methodology, Software}

\affiliation[1]{organization={University of Leicester},
    city={Leicester},
    country={UK}}

\begin{abstract}
In this paper, we study the single-item economic lot-sizing problem with production cost functions that are piecewise linear.
The lot-sizing problem stands as a foundational cornerstone within the domain of lot-sizing problems. It is also applicable to a variety of important production planning problems which are special cases to it according to \cite{ou}.
The problem becomes intractable when $m$, the number of different breakpoints of the production-cost function is variable as the problem was proven NP-hard by \cite{Florian1980}. For a fixed $m$ an $O(T^{2m+3})$ time algorithm was given by \cite{Koca2014} which was subsequently improved to $O(T^{m+2}\log(T))$ time by \cite{ou} where $T$ is the number of periods in the planning horizon.\newline
We introduce a more efficient $O(T^{m+2})$ time algorithm for this problem which improves upon the previous state-of-the-art algorithm by Ou and which is derived using several novel algorithmic techniques that may be of independent interest.

\end{abstract}


 \maketitle

\section{Introduction}
The single-item economic lot-sizing problem where production cost functions are piecewise linear (ELS-PL) is defined as follows:
Given $T$ time-periods in the planning horizon, the demand $d_j$ for each period $0<j\leq T$, an inventory cost function $H_j$ that is general concave over intervals $(-\infty,0]$ and $[0,\infty)$ with $H_j(0)=0$ and a production cost function $P_j$ that is a piece-wise linear function with $m$ different, stationary breakpoints denoted as $B_i$ for $0<i\leq m$, the problem asks for the optimal choices of the production quantity $X_j$ for each period to minimize the total cost assuming that backlogging is allowed and given that the remaining inventory $I_j$ for each period incurs a cost $H_j(I_j)$.
For simplicity, we assume that $B_0=0$.
Furthermore, we specifically define $P_j(0)=0$ and $P_j(X_j)=s_{j,\ell}+p_{j,\ell}X_j$ if $B_{\ell-1}<X_j\leq B_\ell$ for $1\leq \ell\leq m+1$ where $s_{j,\ell}$ and $p_{j,\ell}$ represent the corresponding fixed setup cost and the variable unit production cost respectively for $X_j$ within the interval $(B_{\ell-1}, B_{\ell}]$.

More formally the problem is described as a mathematical program as follows:
\begin{equation*}
\begin{array}{ll@{}lll}
\text{min}  & \displaystyle\sum\limits_{j=1}^{T} (P_j(X_j)+H_j(I_j)) & \, & ( 1) &\\
\text{subject to}& I_j=I_{j-1}+X_j-d_j  \textrm{ where } j=1,2,..T  &\, & ( 2) &\\ &I_0=I_N=0  &\, &( 3) &\\ &X_{j} \geq 0  \textrm{ where } j=1,2,..T &\,& ( 4)
\end{array}
\end{equation*}
The Objective function (1) seeks to minimize the total production and inventory costs.
Constraints (2), (3) and (4) describe the inventory balance equation, the boundary conditions of the function $I$, and the restriction to non-negativity of $X_j$ respectively.
An application that demonstrates the significant practical value of the ELS-PL is given towards the end of the introduction of \cite{ou}.
Besides our main result which is the state-of-art algorithm for the ELS-PL several of the methods used may be of independent interest.
Particularly the Algorithm of Lemma 3 which solves efficiently a sorting problem related to the weighted combinations of sets of elements with weights and the Algorithm of Theorem Y which can be utilized to solve efficiently a class of dynamic programs whose recurrences follow a structure that we describe in the last section.

\section{Literature Review}
The basic ELS problem and some of its variants were formally introduced by \cite{Wagner1958}.
Since the production cost functions of ELS-PL have a very general structure it is considered an important variation of ELS. This generality of ELS-PL also makes many other variations of the ELS problem easily reducible to it (\cite{ou}).

Some of the earliest studied ELS problems with general cost functions were studied by \cite{Florian1971} who provided an optimal $O(\hat{D}\cdot\hat{C})$ time algorithm for the problem with the most general production cost structure, where $\hat{D}$ is the total demand and $\hat{C}$ is the total production capacity of the planning
horizon.

An $O(T\cdot\hat{D}\cdot\hat{m})$ exact algorithm for the case with piecewise linear production costs and general inventory holding costs, where $\hat{m}$ is the average number of breakpoints of the production cost functions was developed by \cite{Shaw1998}.
 
Numerous researchers have delved into the study of ELS models incorporating piecewise concave or linear production/inventory costs, including \cite{Love1973}, \cite{Swoveland1975}, \cite{Baker1978}, \cite{Leung1989} amidst others.
All the aforementioned papers contain algorithms for these ELS models that are either pseudo-polynomial or exponential.

Another relevant problem is the capacitated economic lot-sizing (CELS) problem for which the production capacity of each period is assumed to be different, which implies that each production cost function has a single different breakpoint, which is NP-hard \cite{Florian1980}.
The ELS with constant capacity and concave production/inventory costs (ELS-CC) can be solved in polynomial time \cite{Florian1980},\cite{ou}.
The ELS problem where the production costs are piecewise concave and the number
of breakpoints is fixed is also solvable in polynomial time \cite{Koca2014}.

Other relevant developments in the inventory-capacitated lot-sizing problems (ICLSP) are the discovery of a linear algorithm for the special case of ICLSP where there are zero setup and inventory holding costs by \cite{Lin} where they also coin the name vehicle refueling, after its main application, for this problem.
A generalization of the vehicle refueling problem on general graphs is provided by \cite{Khuller} who solve the problem in polynomial time and also provide approximate solutions to several other NP-complete variants of the problem.

The economic lot-sizing model with piecewise concave production costs (ELS-PC) with fixed $m$ was studied in \cite{Koca2014} which gives an $O(T^{2m+3})$ time algorithm for this problem. 
The ELS-PL is a special case of ELS-PC, where production
costs are assumed to be piecewise linear instead of piecewise concave.
\cite{ou} provided an  $O(T^{m+2}\log(T))$ time algorithm for the ELS-PL problem by a novel approach that uses the RMQ (Range minimum Query) data structure.
In this paper, we provide a more efficient $O(T^{m+2})$ time algorithm for the ELS-PL that makes use of several novel algorithmic methods which may be of independent interest.
According to \cite{ou} "lots of ELS models in the lot-sizing literature are special cases of our model. For example, ELS with minimum ordering quantity (see \cite{Hellion2012},: \cite{Okhrin2011}, among others), ELS with capacity reservation (see \cite{Lee2013}, among others), ELS with all-units discount (see \cite{Li2012}, among others), ELS with subcontracting (see \cite{Lee1989}, among others), ELS with multi-mode replenishment (see \cite{JARUPHONGSA2005}, among others), and so on...  our exact algorithm is capable of improving upon the running time complexity of solving those lot-sizing problems."
As was already noted by \cite{ou}  ELS-PL algorithms can be used to improve the time complexity of a great variety of previously state of art algorithms of existing ELS problems and as such any improved ELS-PL algorithm may constitute a significant contribution to the ELS literature.

\section{Overview of the Presentation}
Before introducing our contributions, we will first present Ou's ELS-PC algorithm, which is a simplified version of the algorithm proposed by \cite{Koca2014} for the same problem, both exhibiting identical time complexity. 
Furthermore, we will introduce several definitions and relevant lemmas from \cite{ou} .
Subsequently, we will unveil our $O(T^{m+2)})$ time algorithm, stemming from the aforementioned algorithm.

\section{An $O(T^{2m+3})$ time Algorithm for the ELS-PC}

Before proceeding to present the algorithm in detail we will give some definitions and a lemma that is required in the derivation of the algorithm for the ELS-PC.

A  regeneration period $r$ is a production period where $I_{r-1}=0$.

Two regeneration periods are consecutive if there isn't any regeneration period between them.
A production period is a period where $X_{j}>0$. 
A fractional period $j$ is a production period where $X_{j} \notin\left\{B_{1}, B_{2}, \ldots, B_{m}\right\}$.

A $B_{\ell}$-period is a period where $X_{j}=B_{\ell}$ for an $\ell\in \{0,1, \ldots, m\}$. 

\begin{lemma}[\cite{Swoveland1975}]
For any instance of ELS-PC there is an optimal solution that for any two consecutive regeneration periods $r$ and $w+1$ where $r\leq w$ there exists at most one fractional period among these production periods 
\end{lemma}

We can now begin describing Ou's algorithm. The algorithm makes use of several recursive functions the main of which is $\Psi_{u}$ which is defined as the optimal total cost satisfying the demand in periods $u, u+1, \ldots, T$, where period $u$ is a regeneration period.

$\psi_{u, v}$ is defined as the minimal cost induced in periods $u, u+1, \ldots, v$, assuming that $u$ and $v+1$ are regeneration periods and that the number of fractional periods in $\{u, u+1, \ldots, v\}$ is at most one. 
The following method, $\mathbf{D P}_{1}$ solves problem ELS-PL optimally with the use of lemma 1:
\newline
(1) Recurrence relation: For $u=1,2, \ldots, T$,

\begin{equation*}  
\tag{5}
\Psi_{u}=\min _{v=u, u+1, \ldots, T}\left\{\psi_{u, v}+\Psi_{v+1}\right\}
\end{equation*}
(2) Boundary condition: $\Psi_{T+1}=0$.
\newline
(3) Objective: $\Psi_{1}$.
\newline

The time complexity of $\mathbf{D P _ { 1 }}$ is $O\left(T^{2}\right)$ assuming the totality of $\psi_{u, v}$ values are given. 

Before we proceed with the description of the rest of the algorithm we will give a few more definitions:

$D(i, j)=\sum_{k=i}^{j} d_{k}$ denotes the cumulative demand for periods $i, i+1, \ldots, j$ (if $i>j$ then $D(i, j)=0$). 

$\Pi$ is the set of integer vectors $\mathbf{n}=\left(n_{0}, n_{1}, \ldots, n_{m}\right)$ that satisfy $0 \leq n_{\tau} \leq T$ for $\tau=$ $0,1, \ldots, m$. 

For any $\mathbf{n}=\left(n_{0}, n_{1}, \ldots, n_{m}\right) \in \Pi$, we define $\nu(\mathbf{n})=\sum_{\tau=0}^{m} n_{\tau}$ and $\omega(\mathbf{n})=\sum_{\tau=0}^{m} n_{\tau} \cdot B_{\tau}$.
For $k=0,1, \ldots, T$, denote $\Pi_{k}=\{\mathbf{n}\in\Pi\mid$ $v(\mathbf{n})=k\}$. 
The vector $\mathbf{n}$ represents the production arrangement for $\nu(\mathbf{n})$ consecutive time periods in which exactly $n_{\tau}$ periods have a production quantity of $B_{\tau}(\tau=0,1, \ldots, m)$, whereas $\omega(\mathbf{n})$ represents the total quantity of production that occurred during the $n_{\tau}$ consecutive periods.

For a vector $\mathbf{n}$ the function $\nu(\mathbf{n})$ represents the number of consecutive time periods, and $\omega(\mathbf{n})$ represents the total production quantity in the $\nu(\mathbf{n})$ consecutive time periods.

For any $\mathbf{n}=\left(n_{0}, n_{1}, \ldots, n_{m}\right)$ and $\mathbf{N}=\left(N_{0}, N_{1},\ldots, N_{m}\right)$ in $\Pi$,  $\mathbf{n} \leq \mathbf{N}$ is defined as $n_{\tau} \leq N_{\tau}$ for $\tau=0,1, \ldots, m$.
We also define the subtraction of the two vectors as follows:
$\mathbf{N}-\mathbf{n}=$ $\left(N_{0}-n_{0}, N_{1}-n_{1}, \ldots, N_{m}-n_{m}\right)$. 
For $\ell=1,2, \ldots, m+1$,  $\mathbf{e}_{\ell} \in \Pi$ is defined as the vector in which the $\ell$th element is equal to 1 and all of the other elements are equal to 0 . \newline
We also define the function $W_\ell(\mathbf{N})$ for any $0\leq i\leq m$ as the function that gives the $\ell$th element of vector  $\mathbf{N} \in \Pi$ . 
To avoid confusion we will use the above function to specify components of vectors and retain the simple index notation $\mathbf{N}_{\ell} \in \Pi$ to specify different vectors.
Let $P_{j}\left(X_{j}\right)=+\infty$ if $X_{j}<0$ or $X_{j}>B_{m+1}$ for $j=1,2, \ldots, T$.

We define \newline  $\Pi_{i-u+1}=$ $\left\{\left(n_{0}, n_{1}, \ldots, n_{m}\right) \in \Pi \mid \sum_{\tau=0}^{m} n_{\tau}=i-u+1\right\}$.
The set can be informally defined as all the vectors of length $m$ whose elements sum to the specific value of $i-u+1$. 
The usefulness of this definition lies mainly in the fact that the set of $\Pi_{i-u+1}$ contains all the vectors of possible breakpoint combinations that must happen in the range of $i-u+1$ consecutive periods without the specificity of the order that they could happen. This way we can use dynamic programming to find the total minimal cost for any $\mathbf{n}=\left(n_{0}, n_{1}, \ldots, n_{m}\right) \in \Pi_{i-u+1}$ that can be applied on a specific range of  consecutive time-periods so that there are exactly $n_{\ell}$ periods each of which has a production quantity of $B_{\tau}$ for $\tau=0,1, \ldots, m$ .
A high-level description of how $\psi_{u, v}$ is computed is as follows:
For each possible fractional period $i$ with  $1 \leq u\leq i \leq v \leq T$  we compute the total minimal cost for the periods between $u, u+1, \ldots, i-1$  and $i+1, i+2, \ldots, v$ using the functions $\bar{f}_{u, i}(\mathbf{n})$ and $\hat{f}_{j, v}(\mathbf{N})$  with $\mathbf{n} \in \Pi_{i-u+1}$ and $\mathbf{N} \in \Pi_{i-v+1}$ respectively.  
$\psi_{u, v}$ is the minimum value among the optimal values computed for each fixed fractional period $i$.
The main observation that allows us to compute  $\bar{f}_{u, i}(\mathbf{n})$ efficiently is that for any $i$ the ending inventory level of period $i$ must be equal to $\sum_{\tau=0}^{m} n_{\tau}$. $B_{\tau}-\sum_{j=u}^{i} d_{j}=\omega(\mathbf{n})-D(u, i)$ since the cost of the ending inventory level, $H_{i}(\omega(\mathbf{n})-D(u, i))$  can be calculated without burdening the complexity of any of the sub-problems that are created.
Thus Ou defines the minimal cost function   $\bar{f}_{u, i}(\mathbf{n})$ for the periods $u,..,i$  by the following recursive equation:

For any $1 \leq u \leq i \leq T$ and $\mathbf{n} \in \Pi_{i-u+1}$,

\begin{equation*}\tag{6} \begin{split}\bar{f}_{u, i}(\mathbf{n})=\min _{\substack{\tau =0,1, \ldots, m \\ \text {s.t.} e_{\tau+1} \leq \mathbf{n}}}\{\bar{f}_{u, i-1}(\mathbf{n}-\mathbf{e}_{\tau+1})
+P_{i}(B_{\tau})+H_{i}(\omega(\mathbf{n})-D(u, i))\}\end{split}\end{equation*}

with the following boundary conditions:
\begin{equation*}\tag{7}\begin{split}\bar{f}_{u, u-1}(n_{0}, n_{1}, \ldots, n_{m})=\begin{cases}0, & \text {if }(n_{0},n_{1}, \ldots,n_{m})=(0,0,\ldots,0) ; \\ +\infty, & \text {otherwise }\end{cases}\end{split}\end{equation*}

Analogously $\hat{f}_{j, v}(\mathbf{N})$ is the minimal total cost for periods $j, j+1, \ldots, v$, given that $v+1$ is a regeneration period.
Furthermore, in $\hat{f}_{j, v}(\mathbf{N})$, the ending inventory level of period $j$ must be equal to $D(j+1, v)-\omega\left(\mathbf{N}-\mathbf{e}_{\tau+1}\right)$ assuming the production quantity in period $j$ is  $B_{\tau}$.  
Its formula is defined as follows:
 For any $1 \leq j \leq v \leq T$ and $\mathbf{N} \in \Pi_{v-j+1}$,

\begin{equation*}\tag{8}
\hat{f}_{j, v}(\mathbf{N})=\min _{\substack{\tau=0,1, \ldots, m \\
\text { s.t. } e_{\tau+1} \leq \mathbf{N}}}\{\hat{f}_{j+1, v}(\mathbf{N}-\mathbf{e}_{\tau+1})+P_{j}(B_{\tau})+H_{j}(D(j+1, v)-
    \omega(\mathbf{N}-\mathbf{e}_{\tau+1}))\}
\end{equation*}

\begin{equation*} \tag{9}
  {f}_{v+1, v}(N_{0}, N_{1}, \ldots, N_{m})
 = \begin{cases}0, & \text { if }(N_{0}, N_{1}, \ldots, N_{m})=(0,0, \ldots, 0) ; \\
       +\infty, & \text { otherwise. }\end{cases}
\end{equation*}

With the Boundary condition:

\begin{equation*}\tag{10}\begin{split}
\hat{f}_{v+1, v}(N_{0}, N_{1}, \ldots, N_{m})= \begin{cases}0, & \text { if }(N_{0}, N_{1}, \ldots, N_{m})=(0,0, \ldots, 0) ; \\ +\infty, & \text { otherwise. }\end{cases}\end{split}\end{equation*}

Both $\bar{f}_{u, i}(\mathbf{n})$ and $\hat{f}_{j, v}(\mathbf{N})$  can be computed in $O(T^{m+2})$  since each of the possible states for these functions is bounded by $O(T^{2}T^{m})$ with $T^2$ being the upper bound on the possible pairs of starting-ending time periods and $T^m$ being the number of  different vectors of production arrangements that can be used as their arguments. Also, the above functions require $O(m)$ time to compute the optimal value for each of these states but since $m$ is considered a constant in the formulation of the problem it is discarded in our bounds.

Now that we have defined all the component functions required for the function, $\psi_{u, v}$ a final function  $\phi_{u, t, v}(\mathbf{n} ; \mathbf{N})$, which gives the minimal total cost assuming that $u,v$ are two consecutive regeneration periods, $t$ is a fractional period and $\mathbf{n},\mathbf{N}$ are the production arrangements for the periods $u..t-1$ and $t+1..v$ respectively. The production for the fractional period can be calculated by the formula $P_{t}(D(u, v)-\omega(\mathbf{n})-\omega(\mathbf{N}))$  since the demand of the periods $u..v$ minus the production of the periods $u..t-1,t+1..v$ given by the production arrangements $\omega(\mathbf{n}),\omega(\mathbf{N})$ minus the production of the fractional period must be equal to zero.
So the formula is defined as follows:

\begin{equation*}\phi_{u, t, v}(\mathbf{n} ; \mathbf{N})=\bar{f}_{u, t-1}(\mathbf{n})+P_{t}(D(u, v)-\omega(\mathbf{n})-\omega(\mathbf{N}))+
\end{equation*}
\begin{equation*}
H_{t}(D(t+1, v)-\omega(\mathbf{N}))+\hat{f}_{t+1, v}(\mathbf{N})\tag{11}
\end{equation*}

By Lemma 1 and the above equation, for any $1 \leq u \leq v \leq T$, we have

\begin{equation*}
\tag{12}\psi_{u, v}=\min _{t=u, u+1, \ldots, v} \min _{\mathbf{n} \in \Pi_{t-u}} \min _{\mathbf{N} \in \Pi_{v-t}}\left\{\phi_{u, t, v}(\mathbf{n} ; \mathbf{N})\right\}.\end{equation*}

The functions $\hat{f}_{j, v}(\mathbf{N})$ , $\bar{f}_{u, i}(\mathbf{n})$  can be pre-calculated and used in $O(1)$ for specific arguments.
Given the values of  $\hat{f}_{j, v}(\mathbf{N})$  and  $\bar{f}_{u, i}(\mathbf{n})$  the time complexity of  $\phi_{u, t, v}(\mathbf{n} ; \mathbf{N})$ is $O(T^{2m})$ for fixed $u,t,v$ since there are $O(T^m)$ different vectors $n$ and $O(T^m)$ different vectors $N$ which it can take as its two arguments making the total number of possible argument combinations $O(T^{m}T^{m})=O(T^{2m})$ .
The function $\psi_{u, v}$ for fixed $u,v$ requires $O(T^{2m+1})$ time given $\phi_{u, t, v}(\mathbf{n} ; \mathbf{N})$ since the possible combinations of arguments is $O(T^{m}T^{m}T)=O(T^{2m+1})$ , the last $T$ being for the evaluation of every possible fractional period position.
For the $O(T^2)$ possible combinations of $1\leq u\leq v\leq T$ the time complexity of  $\psi_{u, v}$ becomes $O(T^{m}T^{m}TT^2)=O(T^{2m+3})$ which is the total time complexity of the algorithm.

\section{An $O(T^{m+2})$ time algorithm for the ELS-PL}
In order to describe our result we will first very briefly describe the $O(T^{m+2}log(T))$ algorithm by Ou and subsequently define additional terminology and formulas from Ou's work.
By employing a lemma provided below, Ou's algorithm can swiftly identify a portion of the dynamic programming solution, eliminating the need to exhaustively examine every attainable state of the DP formulation for a specified set of parameters. 

One ingredient in Ou's algorithm is the function $\varphi_{u, t}(\mathbf{n})$ used in the computation of the optimal cost of the time periods starting from the regeneration period $u$ and ending at $T$ given production arrangement $\mathbf{n} \in \Pi_{t-u}$ with the condition that $\omega(\mathbf{n})<D(u, T)$.
This formula is given below: 

\begin{equation*}
    \tag{13}\begin{aligned}\varphi_{u, t}(\mathbf{n})=\min _{v=t, t+1, \ldots, T} \min _{\mathbf{N} \in \Pi_{v-t}}\left\{\phi_{u, t, v}(\mathbf{n} ; \mathbf{N})+\Psi_{v+1}\right\}\end{aligned}\end{equation*}.

Given a regeneration period $u$, a fractional period $t$, and a production arrangement $\mathbf{n} = \{n_0, n_1, \ldots, n_m\}$,  the function $\varphi_{u, t}(\mathbf{n})$ represents the minimal total cost incurred in all periods from $u$  to $T$ and as such it can be used in the computation of $\Psi_{u}$ as follows:
\begin{equation*}
\tag{14}
\begin{aligned}
\Psi_{u} & =\min _{u \leq t \leq T} \min _{\substack{\mathbf{n} \in \Pi_{t-u} \\
w(\mathbf{n})<D(u, T)}} \min _{t \leq v \leq T} \min _{\mathbf{N} \in \Pi_{v-t}}\left\{\phi_{u, t, v}(\mathbf{n} ; \mathbf{N})+\Psi_{v+1}\right\} \\
& =\min _{u \leq t \leq T} \min _{\substack{\mathbf{n} \in \Pi_{t-u} \\
w(\mathbf{n})<D(u, T)}}\left\{\varphi_{u, t}(\mathbf{n})\right\} .
\end{aligned}
\end{equation*}
Since  the number of vectors in $\Pi_{t-u}$ is no more than $T^{m}$  $,\Psi_{u}$ can be computed from the previous formula in $O(T^{m+1})$ time given the values of $\varphi_{u, t}$.

For our algorithm, we will also employ the following formula:
\begin{equation*}\tag{15} \varphi_{u,t'}=\min _{t' \leq t \leq T} \min _{\substack{\mathbf{n} \in \Pi_{t-u} \\
w(\mathbf{n})<D(u, T)}} \min _{t \leq v \leq T} \min _{\mathbf{N} \in \Pi_{v-t}}\{\phi_{u, t, v}(\mathbf{n} ; \mathbf{N})+\Psi_{v+1}\}=
\end{equation*}
\begin{equation*}  
\min _{t' \leq t \leq T} \min _{\substack{\mathbf{n} \in \Pi_{t-u} \\
w(\mathbf{n})<D(u, T)}}\{\varphi_{u, t}(\mathbf{n})\} .\end{equation*}

Observe that $\Psi_{u}=\varphi_{u,u}$ 
The formula $\varphi_{u, t}$ is used by our algorithm to partially compute the solution of $\Psi_{u}$. 
 
For any $t$,  $\hat{\Pi}_{t}^{\prime}=\{\mathbf{N} \in \Pi \mid \nu(\mathbf{N}) \leq T-t\}=\bigcup_{i=0}^{T-t} \Pi_{i}$, we also define
for any $t$,  $\tilde{\Pi}_{t}^{\prime}=\{\mathbf{n} \in \Pi \mid \nu(\mathbf{n}) \leq t\}=\bigcup_{i=0}^{t} \Pi_{i}$.  

For all  $t \geq 1$ and $\mathbf{N}=\left(N_{0}, N_{1}, \ldots, N_{m}\right) \in \hat{\Pi}_{t}^{\prime}$,  the inventory level at the end of period $t$ given that $I_{t+v(\mathbf{N})}=0$ is defined as follows:

$\hat{I}_{t}(\mathbf{N})=D(t+1, t+\nu(\mathbf{N}))-\omega(\mathbf{N})$

Similarly, for all  $t \geq 1$ and $\mathbf{N}=\left(n_{0}, n_{1}, \ldots, n_{m}\right) \in \tilde{\Pi}_{t}^{\prime}$,  we define

$\tilde{I}_{t}(\mathbf{n})=\omega(\mathbf{n})-D(t-\nu(\mathbf{n}), t-1)$ for any $t\geq\nu(\mathbf{n})$.

Furthermore for any $t<T$ and vector \newline $\mathbf{N}=\left(N_{0}, N_{1}, \ldots, N_{m}\right) \in \hat{\Pi}_{t}^{\prime}$, we define the inventory cost for period $t$ plus the total cost of periods $t+1, t+2, \ldots, T$, with  $I_{t+v(\mathbf{N})}=0$, given that among periods $t+1, \ldots, t+\nu(\mathbf{N})$ there are exactly $N_{\ell}$ periods that have production quantity of $B_{\tau}$ for $\tau=0,1, \ldots, m$  as follows:

\begin{equation*}\tag{16}F_{t}(\mathbf{N})=H_{t}(\hat{I}_{t}(\mathbf{N}))+\hat{f}_{t+1, t+\nu(\mathbf{N})}(\mathbf{N})+\Psi_{t+\nu(\mathbf{N})+1}\end{equation*}

The total cost for the periods $t, t+1, \ldots, T$, given that the inventory cost for period $t$ plus the total cost for periods $t+1, t+2, \ldots, T$ is equal to $F_{t}(\mathbf{N})$, and given the production cost for $t$ is equal to $p_{t, \ell}*\hat{I}_{t}(\mathbf{N})$ is defined by the following formula:

\begin{equation*}
    \tag{17} G_{t, \ell}(\mathbf{N})=p_{t, \ell} \cdot \hat{I}_{t}(\mathbf{N})+F_{t}(\mathbf{N})\end{equation*}
For all  $\ell \in\{1,2, \ldots, m+1\}$ with  $t<T$ and $\mathbf{N} \in \hat{\Pi}_{t}^{\prime}$, 

The production quantity for the factional period $t$, denoted by $X_{t}$ is defined as follows:

$$
\begin{aligned}
X_{t} & =D(u, t+\nu(\mathbf{N}))-\omega(\mathbf{n})-\omega(\mathbf{N}) \\
& =D(u, t)-\omega(\mathbf{n})+[D(t+1, t+\nu(\mathbf{N}))-\omega(\mathbf{N})] \\
& =D(u, t)-\omega(\mathbf{n})+\hat{I}_{t}(\mathbf{N})
\end{aligned}
$$

As it is noted by Ou, $X_{t}$ is within the interval $\left(B_{\ell-1}, B_{\ell}\right]$ for some $\ell \in$ $\{1,2, \ldots, m+1\}$. For any given $\ell, u, t$ and $\mathbf{n} \in \Pi_{t-u}$, the set of possible vectors in $\hat{\Pi}_{t}^{\prime}$ for which  $X_{t} \in\left(B_{\ell-1}, B_{\ell}\right]$ is defined as follows:

\begin{equation*}
\begin{split}
& V_{u, t}^{\ell}(\mathbf{n})=\left\{\mathbf{N} \in \hat{\Pi}_{t}^{\prime} \mid B_{\ell-1}<D(u, t)-\omega(\mathbf{n})+\hat{I}_{t}(\mathbf{N}) \leq B_{\ell}\right\}\end{split}
\end{equation*}
\begin{equation*}
\begin{split}
=\{\mathbf{N} \in \hat{\Pi}_{t}^{\prime} \mid B_{\ell-1}-D(u, t)+\omega(\mathbf{n})<\hat{I}_{t}(\mathbf{N})\\\leq B_{\ell}-
 D(u, t)+\omega(\mathbf{n})\}\end{split}
\end{equation*}
Ou reorders and re-indexes the elements of $\hat{\Pi}_{t}^{\prime}$ so as to create an ordered sequence $\hat{\Pi}_{t}^{\prime}=\left\{\mathbf{N}_{1}, \mathbf{N}_{2}, \ldots, \mathbf{N}_{q}\right\}$ where $\hat{I}_{t}\left(\mathbf{N}_{1}\right) \leq \hat{I}_{t}\left(\mathbf{N}_{2}\right) \leq \cdots \leq \hat{I}_{t}\left(\mathbf{N}_{q}\right)$ with $q$ being the number of vectors of $\hat{\Pi}_{t}^{\prime}$. 

This way, for each $V_{u, t}^{\ell}$ if $V_{u, t}^{\ell}(\mathbf{n}) \neq \emptyset$, there are integers $a$ and $b$ such that $1 \leq a \leq b \leq q$ and

$V_{u, t}^{\ell}(\mathbf{n})=\left\{\mathbf{N}_{a}, \mathbf{N}_{a+1}, \ldots, \mathbf{N}_{b}\right\}$.
For the purposes of our algorithm, we create another ordered sequence in a similar spirit as the above for  $\mathbf{n}\in \tilde{\Pi}_{t}^{\prime}$ by element reindexing such as  $\tilde{\Pi}_{t}^{\prime}=\left\{\mathbf{n}_{1}, \mathbf{n}_{2}, \ldots, \mathbf{n}_{q}\right\}$ where  $\tilde{I}_{t}\left(\mathbf{n}_{1}\right) \leq \tilde{I}_{t}\left(\mathbf{n}_{2}\right) \leq \cdots \leq \tilde{I}_{t}\left(\mathbf{n}_{q}\right)$.

For any $\mathbf{N}, \mathbf{N}^{\prime} \in V_{u, t}^{\ell}(\mathbf{n})$, a vector $\mathbf{N}$ dominates $\mathbf{N}^{\prime}$ if
$\phi_{u, t, t+\nu(\mathbf{N})}(\mathbf{n} ; \mathbf{N})+\Psi_{t+\nu(\mathbf{N})+1} \leq \phi_{u, t, t+\nu\left(\mathbf{N}^{\prime}\right)}\left(\mathbf{n} ; \mathbf{N}^{\prime}\right)+\Psi_{t+\nu\left(\mathbf{N}^{\prime}\right)+1}$.

For any vector $\mathbf{n} \in \Pi_{t-u}$  Ou's algorithm finds the optimal vector $\mathbf{N}$ from each of the sets  $V_{u, t}^{\ell}(\mathbf{n})$ such as that specific vector dominates every other vector in the set, that is $\phi_{u, t, t+v(\mathbf{N})}(\mathbf{n} ; \mathbf{N})+\Psi_{t+v(\mathbf{N})+1} \leq \phi_{u, t, t+v\left(\mathbf{N}^{\prime}\right)}\left(\mathbf{n} ; \mathbf{N}^{\prime}\right)+\Psi_{t+v\left(\mathbf{N}^{\prime}\right)+1}$ for any vector $N'\in V_{u, t}^{\ell}(\mathbf{n})$.

In order to find the vector $\mathbf{N}$ efficiently for each set $V_{u, t}^{\ell}(\mathbf{n})$ both Ou's algorithm and our own method, use the function $G$ which has the property presented in the following lemma.
 
\begin{lemma}[\cite{ou}]
For any $\mathbf{N}, \mathbf{N}^{\prime} \in V_{u, t}^{\ell}(\mathbf{n})$, vector $\mathbf{N}$ dominates $\mathbf{N}^{\prime}$ if $G_{t, \ell}(\mathbf{N})$ $\leq G_{t, \ell}\left(\mathbf{N}^{\prime}\right)$.
\end{lemma}

To provide a clearer understanding of our approach we will give a brief description of our algorithm and how the previously defined concepts and definitions are utilized by it.

Our algorithm combines the reindexing of the elements of the sets $V_{u, t}^{\ell}(\mathbf{n})$, and the lemma by Ou.
It also evaluates each function $\varphi_{u,t}$ using the production arrangement vectors in increasing order of their $\tilde{I}_{t}(\mathbf{n})$  values and uses a monotonicity property of the set to apply a technique that we call monotonic rebordering of the sets $V_{u, t}^{\ell}(\mathbf{n})$ between any two consecutive elements $\mathbf{n_i},\mathbf{n_{i+1}}\in \tilde{\Pi}_{t}^{\prime}$ along with a data structure to keep track of the optimal elements of each of the sets.
For the efficient ordering of the elements in $\hat{\Pi}_{t}^{\prime}$ and $\tilde{\Pi}_{t}^{\prime}$ we use the procedure presented in the proof of the subsequent lemma.

Before we state the lemma we make the following useful observations:
We observe that quantities such as constants added to every element in a tuple don't change the relative order of the tuple's elements.
As such we can freely add such quantities to our tuples of elements without affecting their order.
Furthermore, the sign of each element in a sorted tuple can be changed from positive to negative and vice versa, and then the resulting tuple can be sorted again in linear time.
In the following lemma, we are dealing with the ordering of numbers and not directly with their corresponding vectors. To be able to retrieve the vector that corresponds to a specific number in the tuple we need to store for each element/number its corresponding vector.
Let $S(t)$ denote the tuple of all the  $\omega(\mathbf{N})$ values where $\mathbf{N}\in \hat{\Pi}'_{t}$, that is $S(t)=\{\omega(\mathbf{N})-D(t+1, t+\nu(\mathbf{N}))  \mid \mathbf{N} \in \hat{\Pi}'_{t}\}$ with $\hat{S}(t)$ representing $S(t)$ with its elements in increasing order.  The order of this tuple is the reverse of the order of the same tuple with the omega and the demand functions having their signs reversed, that is $\{-\omega(\mathbf{N})+D(t+1, t+\nu(\mathbf{N}))  \mid \mathbf{N} \in \hat{\Pi}'_{t}\}$. 
\begin{lemma}
For all $t=1,2..T$, the elements of each of the tuples $S(t)$, which can be interpreted as the tuples of all the different weighted sums of combinations of elements with repetitions $\multiset{m+1}{T-t}$+ $\multiset{m+1}{T-t-1}$+$\dots$+ $\multiset{m+1}{1}$ $=O(T^{m+1})$ where each selected element $i\in \{1..m+1\}$ maps into a specific weight $B_i$, can be sorted in $O(T^{m+2})$ total time.
\end{lemma}
To prove the above claim we will provide an algorithm with the required time complexity.
For a fixed $t$ and an arbitrary fixed component at index $j$ of the vectors in $\hat{\Pi}'_{t}$ let $S(t,j)$ represent the tuple of the numbers of the form $\omega(\mathbf{N})-D(t+1, t+v(\mathbf{N}))$ of all the vectors of $\mathbf{N}\in\hat{\Pi}'_{t}$ that have at least one element with index $j$, that is $W_j(\mathbf{N})>0$ and let $\hat{S}(t,j)$ represent the tuple which contains only the values $\omega(\mathbf{N})-D(t+1, t+v(\mathbf{N}))$ of the vectors that have $W_j(\mathbf{N})=0$.

Clearly the elements of $S(t)$ can be created by the union of the elements of $S(t,j)$ and $\hat{S}(t,j)$,
so if we have the last two tuples sorted we can sort $S(t)$ in time linear in the number of the elements of each of the two tuples by merging the two tuples using the Merge procedure from Merge sort. 
Observe that since $\hat{S}(t,j)$ contains only the omega values of vectors with $m$ components thus the tuple can be sorted by a regular sorting algorithm in $O(T^{m}log(T))$ time since there are only $O(T^m)$ vectors.
As for $S(t,j)$ observe that if we subtract $B_j$ and add $D(t+1,t+1)$ from each of its elements it is transformed into $S(t+1)$ where each element of one tuple corresponds to an equivalent one of the other, so if we have $\hat{S}(t+1)$ already we can derive the ordered $S(t,j)$ by adding $B_j$ and subtracting $D(t+1,t+1)$ from each of its elements without affecting the order of the elements.
Observe that for this transformation we don't need to subtract from any element of $\hat{S}(t+1)$ (to convert them to elements of the set $S(t,j)$) the quantity $D(t+v(\mathbf{N}), t+\nu(\mathbf{N}))$ ) since the value of $v(\mathbf{N}) $ for each $\mathbf{N}\in S(t+1)$ is increased by 1 thus offsetting the decrease by 1 of $t$.
Thus, given that $S(T-1)$ can be sorted in $O({m}log(m))$ time since it contains all the omega values of vectors $\mathbf{N}\in \hat{\Pi}_{T-1}$ with a single element, that is $\nu(\mathbf{N})=1$, it can be shown by induction on $t$ that each $S(t)$ can be sorted in linear time. So $S(t)$ for all $0<t\leq T$ can be sorted in total $O(T^{m+2})$ time thus concluding our proof.

The tuples $S(t)$ for all $t=1..T$ can be sorted even faster than $O(T^{m+2})$ by observing that if we keep each $S(t+1)$ starting from $S(T-1)$ in a self-balancing binary tree and retain sums of the values of $B_j$ as an offset for its elements \footnote[1]{To keep track of the updated values of each of the elements in the tree we need to retain the information of the highest value of $y$ for which they appeared in a $\hat{S}(y,j)$ so as to be able to add the relevant multiple of $B_j$ to them when we compare them with each newly inserted element in the tree} then the insertions of each of the $O(T^{m})$ elements of $\hat{S}(t,j)$ is $log(T)$ thus making  $O(T^{m+1}log(T))$ the total time to sort $S(t)$ for all $t$.

The above procedure can be appropriately modified to order the elements of  $\tilde{I}_{t}(\mathbf{n})=\omega(\mathbf{n})+D(t-\nu(n), t)$ with $\mathbf{n}\in \tilde{\Pi}_{t}^{\prime}$ for any $t$.

Now that we have our efficient sorting method we can proceed with a more detailed description of our main method.
Our main algorithm can be synoptically described as follows:

\begin{itemize}
    \item Starting with $t=T$ and going backward $(T,T-1..1)$ for each fixed $t$ we compute $\varphi_{u, t}(\mathbf{n}) $ for every product arrangement $\mathbf{n}\in \tilde{\Pi_t}'$ and every $u\leq t$ in a specific order that enables us to efficiently transform the solution sets of any two successive $\mathbf{n_i},\mathbf{n_{i+1}}\in\tilde{\Pi_{t}}'$ product arrangements from $V_{u, t}^{\ell}(\mathbf{n_i})$ to $V_{u', t}^{\ell}(\mathbf{n_{i+1}})$ for any $\ell$ where $u=t-\nu(n_i)$ and $u'=t-\nu(n_{i+1})$ in $O(1)$ amortized time.
    \item The creation and transformation of the sets $V_{u, t}^{\ell}(\mathbf{n})$ are done efficiently by a method that creates monotonicity in the change of the endpoints of the sets to allow efficient tracking of both the changes that are required for each set as well as the optimal values of the set.
    \item So after having computed all $\varphi_{u, t}(\mathbf{n}) $ for $t$ we have the values of all  $\varphi_{u,t'}$  for $t'\leq t$ and the partial values of  $\varphi_{u,t''}$  for $t\geq t''$.
    \item The sorting of  the $\hat{I}_{t}(\mathbf{N})$ values is done efficiently using the procedure we provided in the proof of our lemma 3.

\end{itemize}

\subsection{Creation, use, and rebordering of the relevant structures}

Before diving into the specifics of the operations performed on the solution sets $V_{u, t}^{\ell}(\mathbf{n})$ with $\mathbf{n}\in\tilde{\Pi_{t}}'$ for a fixed $t$, we will introduce the necessary terminology.
To each set $V_{u, t}^{\ell}(\mathbf{n})$ we associate two linked lists $Q_{u, t}^{\ell}(\mathbf{n})$ and $\hat{Q}_{u, t}^{\ell}(\mathbf{n})$.
Each linked list $Q_{u, t}^{\ell}(\mathbf{n})$ is used to store the optimal $\mathbf{N} \in V_{u, t}^{\ell}(\mathbf{n}) $ vectors in non-decreasing  $G_{t, \ell}(\mathbf{N})$ values while each of the $\hat{Q}_{u, t}^{\ell}(\mathbf{n})$ is are used to store the vectors  $\mathbf{N} \in V_{u, t}^{\ell}(\mathbf{n}) $ in non-decreasing $\hat{I}_{t}(\mathbf{N})$  values e.g $\hat{I}_{t}\left(\mathbf{N}_{1}\right) \leq \hat{I}_{t}\left(\mathbf{N}_{2}\right) \leq \cdots \leq \hat{I}_{t}\left(\mathbf{N}_{q}\right)$.

The borders of a set $V_{u, t}^{\ell}(\mathbf{n})$ are the values $B_{\ell-1}-D(u, t)+\omega(\mathbf{n})$ and $B_{\ell}-D(u, t)+\omega(\mathbf{n})$  from the set description $\{\mathbf{N} \in \hat{\Pi}_{t}^{\prime} \mid B_{\ell-1}-D(u, t)+\omega(\mathbf{n})<\hat{I}_{t}(\mathbf{N}) \leq B_{\ell}-D(u, t)+\omega(\mathbf{n})\}$that delineate the leftmost and rightmost boundaries of the set. 
The borders of any two consecutive sets $V_{u, t}^{\ell}(\mathbf{n})$ and $V_{u, t}^{\ell+1} (\mathbf{n})$  are always non-intersecting, with the boundaries of the later set always starting after the boundaries of the former set end due to the inequality  $B_{0}<B_{1}<\cdots<B_{m}<B_{m+1}$.
Furthermore assuming that $t$ remains constant the borders of these sets are only dependent on the vector $\mathbf{n}$. 
$u$ is also dependent on the value of the production arrangement vector,  $\mathbf{n}$.
The values of $B_{x}$ for any $1\leq x \leq m$ are given in the problem instance and are assumed to be invariant.

Rebordering of the sets  $V_{u, t}^{1}(\mathbf{n_i}), V_{u, t}^{2}$ $(\mathbf{n_i}),\dots,V_{u, t}^{m}(\mathbf{n_i})$ for any $t,m$ refers to the changing of the conditions of the borders of these sets and the transferring of the appropriate elements from each of the sets to accommodate these changes so as to convert them into the sets   $V_{u', t}^{1}(\mathbf{n_j})$,$V_{u', t}^{2}(\mathbf{n_j}),\dots, V_{u', t}^{m}(\mathbf{n_j})$  for any $j\neq i$.
Since these sets are represented by the linked lists $Q_{u, t}^{\ell}(\mathbf{n})$ and $\hat{Q}_{u, t}^{\ell}(\mathbf{n})$ rebordering of the sets entails the conversion of the linked lists from $Q_{u, t}^{\ell}(\mathbf{n_i})$ and $\hat{Q}_{u, t}^{\ell}(\mathbf{n_i})$  to  $Q_{u, t}^{\ell}(\mathbf{n_j})$ and $\hat{Q}_{u, t}^{\ell}(\mathbf{n_j})$ by transference of elements.

Given a sequence of vectors $\mathbf{n_1},\mathbf{n_2},\mathbf{n_3},..\mathbf{n_T}\in {\Pi}$ , the set of sets $V_{u, t}^{1}(\mathbf{n}),$$V_{u, t}^{2}(\mathbf{n})\dots,V_{u, t}^{m}(\mathbf{n})$  with $\mathbf{n}\in {\Pi}$ and a fixed $t$ and $m$ we call monotonic rebordering the rebordering of $V_{u, t}^{1}(\mathbf{n_i}),V_{u, t}^{2}(\mathbf{n_i})..V_{u, t}^{m}(\mathbf{n_i})$  to $V_{u', t}^{1}(\mathbf{n_{j}}),V_{u', t}^{2}(\mathbf{n_{j}}),\dots,V_{u', t}^{m(\mathbf{n_{j}})}$  for each $\mathbf{n_i},\mathbf{n_j}\in {\Pi}$ , $0<i \leq T$  , $1\leq \ell \leq m$ and $j=i+1\leq T$  with the following property:
$B_{\ell-1}-D(u, t)+\omega(\mathbf{n_i})\leq B_{\ell-1}-D(u', t)+\omega(\mathbf{n_{j}})$ and  $B_{\ell}-D(u, t)+\omega(\mathbf{n_i})\leq B_{\ell}-D(u', t)+\omega(\mathbf{n_{j}})$ .

The property of monotonic rebordering is crucial to our algorithm since it enables us to use efficient methods to transfer and trace the optimal elements of the sets quickly.
Observe that if the sets $V_{u, t}^{1}(\mathbf{n}),V_{u, t}^{2}(\mathbf{n})..V_{u, t}^{m}(\mathbf{n})$ with $\mathbf{n}\in \tilde{\Pi_t}'$  are such that their elements $\mathbf{N}\in \hat{\Pi_t}'$ obey the inequality   $\hat{I}_{t}\left(\mathbf{N}_{1}\right) \leq \hat{I}_{t}\left(\mathbf{N}_{2}\right) \leq \cdots \leq \hat{I}_{t}\left(\mathbf{N}_{q}\right)$ then the rebordering for the sets  $V_{u, t}^{1}(\mathbf{n}),V_{u, t}^{2}(\mathbf{n})..V_{u, t}^{m}(\mathbf{n})$  for the sequence of vectors $\mathbf{n_1},\mathbf{n_2},\mathbf{n_3},..\mathbf{n_q}\in \tilde{\Pi_t}'$  is monotonic since for any two consecutive vectors $\mathbf{n_i},\mathbf{n_{i+1}}\in \tilde{\Pi_t}'$ the monotonic property of the borders for each of the sets holds due to the following inequalities $\tilde{I}_{t}\left(\mathbf{n}_{1}\right) \leq \tilde{I}_{t}\left(\mathbf{n}_{2}\right) \leq \cdots \leq \tilde{I}_{t}\left(\mathbf{n}_{q}\right)$ .
Due to $\tilde{I}_{t}(\mathbf{n}_{i})=\omega(\mathbf{n_i})-D(u, t)$ we observe that the following inequalities of the borders of these sets hold since they can be written as 
$B_{\ell-1}-D(u, t)+\omega(\mathbf{n_i})=B_{\ell-1}+\tilde{I}_{t}(\mathbf{n}_{i})\leq B_{\ell-1}-D(u', t)+\omega(\mathbf{n_{i+1}})=B_{\ell-1}+\tilde{I}_{t}(\mathbf{n}_{i_{i+1}})$ for the leftmost borders and 
$B_{\ell}-D(u, t)+\omega(\mathbf{n_i})=B_{\ell}+\tilde{I}_{t}(\mathbf{n}_{i})\leq B_{\ell}-D(u', t)+\omega(\mathbf{n_{i+1}})=B_{\ell}+\tilde{I}_{t}(\mathbf{n}_{i_{i+1}})$
for the rightmost borders of the sets.

Now that we have established the monotonic property of the sets $V_{u, t}^{1}(\mathbf{n}),V_{u, t}^{2}(\mathbf{n}),\dots,V_{u, t}^{m}(\mathbf{n})$  for the sequence of vectors $\mathbf{n_1},\mathbf{n_2},\mathbf{n_3},\dots,\mathbf{n_q}\in \tilde{\Pi_t}'$ we can proceed to make more observations about the nature of the elements that need to be transferred between the sets for each consecutive vectors $\mathbf{n_i},\mathbf{n_{i+1}}\in \tilde{\Pi_t}'$ and how these can be utilized by the linked list data structures in order to be able to trace efficiently these elements and the optimal values of $G_{t, \ell}(\mathbf{N})$ that are associated with the $\mathbf{N}\in V_{u, t}^{\ell}(\mathbf{n})$ vectors of each of the sets.
As was already described each of these sets $V_{u, t}^{1}(\mathbf{n}), V_{u, t}^{2}(\mathbf{n}),\dots,V_{u, t}^{m}(\mathbf{n})$  contains a subset of the elements $N_1,N_2,N_3,N_4,N_5,..,N_q$ with $N_i\in \hat{\Pi}_{t}^{\prime}$  for  $0< i\leq T$  and since these elements are ordered in non-decreasing order of $\hat{I}_{t}(\mathbf{N}_{i})$ values, each two $V_{u, t}^{i}(\mathbf{n})$  and $V_{u, t}^{j}(\mathbf{n})$ for any  $i\leq j$  contain different elements from $\hat{\Pi}_{t}^{\prime}$ and each of the elements contained in the first set have strictly lower indexes than the elements of the second set.
Also if each of the sets is represented by structures that preserve the non-decreasing order in terms of $\hat{I}_{t}(\mathbf{N}_{i})$ of their elements then for any two consecutive elements $n_i,n_{i+1}$, the same structures storing the elements for $V_{u, t}^{\ell}(\mathbf{n_i})$ for any $1 \leq \ell\leq m$  can be used to store the appropriate elements of  $V_{u', t}^{\ell}(\mathbf{n_{i+1}})$  for the same  $t,\ell$.
The elements contained in the structures representing the sets $V_{u, t}^{\ell}(\mathbf{n_i})$ for any $1 \leq \ell\leq m$ that need to be transferred to a $V_{u', t}^{\ell'}(\mathbf{n_{i+1}})$  for some $1 \leq \ell'\leq m$ with $\ell'\neq\ell$ always form sequences of consecutive elements (a sequence of the elements to it and one for those transferred from it to another set) if they exist at all. 
We also note that each element $\mathbf{N_j}\in V_{u, t}^{\ell} (\mathbf{n_i})$  for any $0< j\leq T$ can only go to a 'lower' set, that is a set $V_{u', t}^{\ell'} (\mathbf{n_{i+1}})$ with $\ell' \leq \ell$  for any  $\mathbf{n_i},\mathbf{n_{i+1}}\in \tilde{\Pi_t}'$ since the boundaries of the sets increase monotonically. So the elements of the sequence $\mathbf{n_1},\mathbf{n_2},\mathbf{n_3},..,\mathbf{n_q}\in \tilde{\Pi_t}'$ can move at most $O(m)$ times to a new set which is $O(1)$ since $m$, is considered a constant due to:

\begin{itemize}
    \item Their monotonic transference to 'lower' sets
    \item The fact that once they reach $V_{u, t}^{1} (\mathbf{x})$ for some $x\in \tilde{\Pi_t}'$ they can't be moved to another set anymore.

\end{itemize}
 By using a linked list $\hat{Q}_{u, t}^{\ell}(\mathbf{n})$ to store the elements of each set $V_{u, t}^{\ell}(\mathbf{n})$ we can utilize the above properties to move all the relevant elements from and to each of the sets quickly for each consecutive $\mathbf{n_i},\mathbf{n_{i+1}}\in \tilde{\Pi_t}'$ while the non-decreasing order of the elements in terms of $\hat{I}_{t}()$ is maintained in each linked list.
Starting from $\ell=m$ and going backward towards $1$  for any consecutive $\mathbf{n_i},\mathbf{n_{i+1}}\in \tilde{\Pi_t}'$ each $\hat{Q}_{u, t}^{\ell}(\mathbf{n_{i+1}})$ can be created from the previous $\hat{Q}_{u, t}^{\ell}(\mathbf{n_{i}})$ by removing from its front each of the elements  $\mathbf{N_j}\in V_{u, t}^{\ell} (\mathbf{n_i})$ that have lower value  $\hat{I}_{t}(\mathbf{N_j})$ than the leftmost border of $\hat{Q}_{u, t}^{\ell}(\mathbf{n_{i+1}})$ and transferring them to $\hat{Q}_{u, t}^{\ell-1}(\mathbf{n_{i+1}})$ by connecting them to the end of that linked list in the order they are removed from the previous list.
We will call the above procedure which takes time linear in the elements removed the set maintenance procedure.

Since the order of the elements within each linked list does not by itself give their optimality with respect to the function 
$G_{t, \ell}(\mathbf{N})=p_{t, \ell} \cdot \hat{I}_{t}(\mathbf{N})+F_{t}(\mathbf{N})$ 
where   $\ell \in\{1,2, \ldots, m+1\}$, $t<T$ and $\mathbf{N} \in \hat{\Pi}_{t}^{\prime}$, we associate a second linked list  $Q_{u, t}^{\ell}(\mathbf{n})$ to each set $V_{u, t}^{\ell}(\mathbf{n})$ .
To maintain the optimal vectors $\mathbf{N}\in V_{u, t}^{\ell}(\mathbf{n_{i+1}})$ in non-decreasing order of their $G_{t, \ell}(\mathbf{N})$ values in the linked lists $Q_{u, t}^{\ell}(\mathbf{n_{i+1}})$ we will use the elements of the associated  $\hat{Q}_{u, t}^{\ell}(\mathbf{n_{i+1}})$ linked lists but before we explain that we make the following relevant observation:
If an element $\mathbf{N_j}\in V_{u, t}^{\ell}(\mathbf{n_{i+1}})$ within a linked list $Q_{u, t}^{\ell}(\mathbf{n_{i+1}})$ for any $\ell,u,t$ and $\mathbf{n_{i+1}}$ has a $G_{t, \ell}(\mathbf{N_j})>G_{t, \ell}(\mathbf{N_d})$ with $\mathbf{N_d}\in V_{u, t}^{\ell}(\mathbf{n_{{i+1}}})$ but the position of $\mathbf{N_d}$ is closer to the rear of the linked list than $\mathbf{N_j}$ , that is $\hat{I}_{t}(\mathbf{N_j})>\hat{I}_{t}(\mathbf{N_d})$ then the vector $\mathbf{N_j}$ will always be suboptimal in the set with respect to the function $G_{t, \ell}()$ for any $\mathbf{n_i},\mathbf{n_{i+1}}\in \tilde{\Pi}'$ since the vector $\mathbf{N_d}$ can only be removed and transferred to another set from the set  $V_{u, t}^{\ell}(\mathbf{n_{i+1}})$ after the vector $\mathbf{N_j}$ has been already transferred so for the duration of  $\mathbf{N_j}$'s existence in that set $\mathbf{N_d}$ will also be in the same set.

So for a fixed $t$ and any transition  $\mathbf{n_i},\mathbf{n_{i+1}}\in \tilde{\Pi_t}'$  as soon as an element $\mathbf{N_j}\in V_{u, t}^{\ell}(\mathbf{n_{i+1}})$ is transferred to a list $\hat{Q}_{u, t}^{\ell}(\mathbf{n_{i+1}})$ we calculate its value $G_{t, \ell}(\mathbf{N_j})$ and starting from the element $N_l$ at the end of that list  $Q_{u, t}^{\ell}(\mathbf{n_{i+1}})$  we check if $G_{t, \ell}(\mathbf{N_j})<G_{t, \ell}(\mathbf{N_l})$ and if the inequality holds we remove the element $\mathbf{N_l}$ and repeat this removal procedure with the next element in the list and $\mathbf{N_j}$. When an element $\mathbf{N_w}\in V_{u, t}^{\ell}(\mathbf{n_{i+1}})$ is encountered where $G_{t, \ell}(\mathbf{N_j})>G_{t, \ell}(\mathbf{N_w})$ we insert $\mathbf{N_j}$ at the end list behind that element and stop.
We will call this removal procedure which is done along with the insertion of an element $\mathbf{N_j}\in V_{u, t}^{\ell}(\mathbf{n_{i+1}})$  to a list $\hat{Q}_{u, t}^{\ell}(\mathbf{n_{i+1}})$ the check-and-remove procedure.
It can be easily proven by induction that for each $\mathbf{n_i}\in \tilde{\Pi_t}'$  the elements contained in the lists $Q_{u, t}^{\ell}(\mathbf{n_{i}})$ are always in non-decreasing order after the above operation is done. 
One last thing to be considered is how the removed elements from each of the lists $\hat{Q}_{u, t}^{\ell}(\mathbf{n_{i}})$ are traced and removed from the associated lists $Q_{u, t}^{\ell}(\mathbf{n_{i}})$. 
This can be done by associating each element $ \mathbf{N}\in \hat{\Pi}$  with a function represented by an array that keeps the index of the current list that contains the element so if it appears in another list it can be skipped in that list, alternatively, such elements can be removed if the lists are doubly linked and for each element in the $\hat{Q}_{u, t}^{\ell}(\mathbf{n_{i}})$ lists a pointer to the same element in the list $Q_{u, t}^{\ell}(\mathbf{n_{i}})$ is kept so it can be removed from the current list if it is transferred into another.
Now that we have defined the necessary structures and operations needed to maintain the relevant data for the efficient operation of our algorithm we can describe how they are used in the efficient calculation of  $\Psi_{u}$ for each $1 \leq u\leq T$.
For a fixed $t$ we can evaluate the following equation for the ordered sequence of production arrangements  $\mathbf{n_i}\in \tilde{\Pi_t}'$ in increasing order of $1\leq i\leq \nu(\tilde{\Pi_t})$ since this order allows the use of the monotonic rebordering of the sets  $V_{u, t}^{1}(\mathbf{n_i}), V_{u, t}^{2}(\mathbf{n_i}),\dots, V_{u, t}^{m}(\mathbf{n_i})$  so we can utilize the data structures that we have defined above:
\begin{equation*}
\tag{18}
  \varphi_{u',t}(\mathbf{n})=\min_{
    \substack{\mathstrut i=1,2,\dots,m \\[0.2ex]
      \mathbf{N}=\argmin\limits_{\mathbf{x}\in V_{u', t}^{i}(\mathbf{n})}
      {G_{t, i}(\mathbf{x})}}}
  \{\phi_{u', t, t+\nu(\mathbf{N})}(\mathbf{n} ;
    \mathbf{N})+\Psi_{t+\nu(\mathbf{N})+1}\}
\end{equation*}
where  $u'=t-\nu(\mathbf{n})$ and $w(\mathbf{n})<D(u,T)$.
Since 

\begin{equation*} \tag{19}
  \begin{aligned}
\Psi_{u}=\min _{u \leq t \leq T} \min _{\substack{\mathbf{n} \in \Pi_{-u}\\
\omega(\mathbf{n})<D(u, T)}} \min _{t \leq v \leq T} \min _{\mathbf{N} \in \Pi_{v-t}}\left\{\phi_{u, t, v}(\mathbf{n} ; \mathbf{N})+\Psi_{v+1}\right\} =\min _{u \leq t \leq T}\left\{\varphi_{u, t}\right\}=\varphi_{u, u}.
\end{aligned}
\end{equation*}

For a fixed $u$ once we have derived $\varphi_{u, t}$ from the computation of $\varphi_{u',t}(\mathbf{n})$ for all $\mathbf{n}\in \vert(\tilde{\Pi_t})'\vert$  and all $T\geq t\ge u$ we can obtain the $\Psi_{u}$ in $O(1)$ time since we can compute equation (19) from (15) that has by then been computed for the starting period $u+1$ from the computation of equation (18) for the relevant values and this holds for any $1\leq u\leq T$. Also observe that since the value of $u'$ in equation (14) is dependent only on  $t$ and $\mathbf{n}$ all the values of $\varphi_{u',t}$ for $t$ and $1\leq u'\leq T$ are also computed and can be reused for the computation of further values of $\Psi_{u''}$ with $ 1\leq u''<u$ .
The equation (18) can be solved for a fixed $1\leq t\leq T$ and all the elements in the sequence $\mathbf{n_i}\in \tilde{\Pi_t}'$ in  $O(\vert\tilde{\Pi_t}'\vert+\vert \hat{\Pi_t}'\vert ) $ time since the operations to find the optimal $\mathbf{N}$ that gives the minimum $G_{t, \ell}(\mathbf{N})$ for each $\ell$ takes constant time and the reconfiguration that is the set maintenance procedure and the removal of suboptimal elements before an element insertion of the lists for all consecutive $\mathbf{n_i},\mathbf{n_{i+1}}\in \tilde{\Pi_t}'$ with $1\leq i<T$  take at most  $O(\vert\tilde{\Pi_t}'\vert+\vert \hat{\Pi_t}'\vert ) $ since this is the total number of elements that can be deleted and inserted and each deleted element is never encountered again in the same list and thus takes no further part in the computation whereas each inserted element can be inserted only once in each of the lists during the whole procedure for a fixed $t$.

\begin{theorem}
For $1\leq u \leq T$, all of the values of $\Psi_{u}$ can be determined in $O\left(T^{m+2}\right)$ total time.
\end{theorem}

\subsection{The complete algorithm}

Prior to the computation of our main algorithm, several prerequisite functions are pre-computed in $O\left(T^{m+2}\right)$ total time. Subsequently, the values of $\Psi_{T}$ are computed backwards starting from $T$ and going to 1.
For the base case when $t=T$  the following formula is used $\Psi_{T}=P_{T}\left(d_{T}\right)$.
For $t=T,T-1,..1$ we compute equation (18) for each of the ordered elements of the sequence  $\tilde{\Pi}_{t}^{\prime}$, compare the values of each computed $\varphi_{u',t}(\mathbf{n})$ with the previous minimum value stored in a temporary array for $\varphi_{u,t}$ and keep the minimum of the two.
$\Psi_{t}$ is then computed using the equation formula (19).
Our algorithm as presented below is for the cases where there is a fractional period $u<t<v$ between two consecutive regeneration periods $u,v+1$. The cases where either the fractional period is at one of the periods $u,v$ or there isn't a fractional period are analyzed in the appendix and are resolved in the same way as in \cite{ou}. 

\section*{Algorithm 1}
\begin{algolist}
\item Compute $\nu(\mathbf{n})$ and $\omega(\mathbf{n})$ for all $\mathbf{n} \in \Pi$, and compute $\bar{f}_{i, j}(\mathbf{n})$ and $\hat{f}_{i, j}(\mathbf{n})$ for all $1 \leq i \leq j \leq T$ and $\mathbf{n} \in \Pi_{j-i+1}$ .
\item We use the procedure of Lemma 3 to generate the sorted vector sets for $t=2,3, \ldots, T-1$ in non-decreasing $\hat{I}_{i}(\mathbf{n})$  values for $\mathbf{n} \in \hat\Pi_{t}'$ and $\tilde{I}_{i}(\mathbf{n})$ values for $\mathbf{n} \in \tilde\Pi_{t}'$.
\item Define  $\Psi_{T}=P\left(d_{T}\right)$. For $u=T-1, T-2, \ldots, 1$ do
    \begin{algolist}
    \item  $M_{u}$ is the minimum of the computed values of all three equations in Appendix B. If $u=T-1$, then let $\Psi_{u}=M_{u}$ and proceed to Step 3.3; otherwise, if $u \leq T-2$, then go to Step 3.2.
    \item Let $t=u+1$ .
        \begin{algolist}
        \item For every $\mathbf{n} \in \tilde\Pi_{t}'$  in increasing $\tilde{I}_{i}(\mathbf{n})$ order with $\omega(\mathbf{n})<D(u, T)$, compute each of the $\varphi_{t-\nu(\mathbf{n}), t}(\mathbf{n})$ values from eq. (18) using the appropriate values of the list ${Q}_{t-\nu(\mathbf{n}) t}^{\ell}(\mathbf{n})$.From each of these values compute $M_{t-\nu(\mathbf{n}),t}=\\\min(M_{t-\nu(\mathbf{n}),t},\varphi_{t-\nu(\mathbf{n}), t}(\mathbf{n}),\\\varphi_{t-\nu(\mathbf{n}), t+1})$.
Subsequently we update lists $\hat{Q}_{t-\nu(\mathbf{n}), t}^{\ell}(\mathbf{n})$ and ${Q}_{t-\nu(\mathbf{n}), t}^{\ell}(\mathbf{n})$ using the set maintenance procedure for the next value of $\mathbf{n} \in \tilde\Pi_{t}'$ and the check-and-remove procedures for any transferred element .
      \\
        
        \item Transfer the value of $M_{u,t}$ to $\varphi_{u, t}$ and compute the $\Psi_{u}$ value via Eq.(19). If  $\Psi_{u}>M_u$ then let  $\Psi_{u}=M_{u}$.
        \end{algolist}

    \item    Add each element $\mathbf{N} \in \Pi_{u}^{\prime}$ to the appropriate lists $\hat{Q}_{t-\nu(\mathbf{n_1}), t}^{\ell}(\mathbf{n_{1}})$ and ${Q}_{t-\nu(\mathbf{n_1}), t}^{\ell}(\mathbf{n_{1}})$ for all $1\leq \ell\leq m+1$, then compute the values of $G_{u, \ell}(\mathbf{N})$ for all $\mathbf{N} \in \Pi_{u}^{\prime}$ and finally starting from the second element of each list ${Q}_{t-\nu(\mathbf{n_1}), t}^{\ell}(\mathbf{n_{1}})$ and going towards the last execute the check-and-remove procedure for each of these elements to remove any element $N_w$ in front of them in the same list that has higher value of $G_{u, \ell}(\mathbf{N_w})$ than itself. 

    \end{algolist}
\end{algolist}

During the first step of our algorithm, we calculate all of the $v(\mathbf{n})$, $\omega(\mathbf{n}), \bar{f}_{u, i}(\mathbf{n})$ and $\hat{f}_{j, v}(\mathbf{N})$ values. 
In Step 2, the vector set $\hat{\Pi}_{i}^{\prime}$ and its sequence are computed along with $\hat{I}_{i}(\mathbf{N})$ for each $\mathbf{N} \in \hat{\Pi}_{i}^{\prime}$. In Step 3, $\Psi_{T}=P\left(d_{T}\right)$ is computed. 
In Step 3.1, for all  $u<T, M_{u}$ is used to calculate the  $\Psi_{u}$ for the aforementioned special cases regarding the position of $t$ which are described in the appendix.

In Step 3.2 the value of $\Psi_{u}$ is calculated for the case where $u<t<v$, this is done with the aid of the linked lists which are used to determine the optimal values in each set and subsequently to efficiently transform the sets for each subsequent element  $\mathbf{n} \in \tilde\Pi_{t}'$ that is evaluated.
In Step 3.3, the values of $G_{u, \ell}(\mathbf{N})$ are calculated and stored in the appropriate lists. After that, the elements of each of the lists ${Q}_{t-\nu(\mathbf{n}), t}^{\ell}(\mathbf{n_{1}})$ are processed to contain only the optimal elements in increasing $G_{u, \ell}(\mathbf{N})$ value.
Step 1 has a time complexity of $O\left(T^{m+2}\right)$. 
For each given $t$,  Step 2 has a complexity of $O\left(T^{m+1}\right)$ if we use the sorting method presented at lemma 3 to precompute the sorted lists of vectors for all relevant $t$ values in  $O\left(T^{m+2}\right)$ total time.
At Step 3.  $\Psi_{T}=P\left(d_{T}\right)$ is computed in $O(1)$ time. For any given $u<T$, the total time complexity of Step 3.1 is $O\left(T^{m+1}\right)$. For any given $u$ and $t$ with $u<t<T$, the complexity of Step 3.2.1 is $O(\vert\tilde{\Pi_t}'\vert+\vert\hat{\Pi_t}'\vert) =O(T^{m+1})$.
Step 3.3 takes $O(\vert\hat{\Pi_t}'\vert) =O(T^{m+1})$ .
The total complexity of Algorithm 1 is $O(TT^{m+1})=O(T^{m+2})$.
We leave several of the more detailed parts of the analysis of the algorithm in the appendix.
\section{Conclusion}
In this paper, we provide an exact $O(T^{m+2})$ time algorithm for the ELS-PL which is asymptotically faster than the previous state-of-art  $O(T^{m+2}\log(T))$algorithm by Ou.
Since the time complexity for several of the components and for the significantly simpler instances of the problem where there are no fractional periods also require $O(T^{m+2})$ time themselves it may be the case that our methods reach the limit in terms of time complexity of what can be achieved with the dynamic programming approach to the problem that was initiated by \cite{Koca2014}.

Our algorithm is also constructed by elementary means which don't require the use of more advanced data structures such as advanced Range minimum query (RMQ) structures as was the case with the previous algorithm.
Of particular note is our sorting method of lemma 3 which divides the elements to be sorted into two sets,  the smallest of which is then sorted on the fly. At the same time, the larger set is sorted using the previously sorted list $S(t)$. Subsequently, both of these sets are combined in time linear of their elements.
Our main method can be used to speed up recursive algorithms that follow the same general structure as described in the previous section and its component methods such as the monotonic rebordering of sets with bordering boundaries may be of independent interest.

\appendix
\section{Appendix A}
\subsection{Proof of Theorem 4}
Here we will provide a more detailed description and analysis of our algorithm 1 and some of its steps.
Firstly we note that the $\Psi_{u}$  values are computed in reverse order starting from $u=T$ and going down to $u=1$. For a fixed $u$, $\Psi_{u}$ is computed using the previously computed values of  $\varphi_{u, t}$.
Note that for a fixed $t$ the parameter $u'$ of the auxiliary array $M_{u',t}$ is defined as $u'=t-\nu(\mathbf{n})$ so several of the $\varphi_{u'', t}$  that are computed aren't used directly in the computation of $\Psi_{u}$ with $u=t-1$ but for the computation of subsequent values of $\Psi_{u''}$ with $u''<u$.
The definite value of each $\varphi_{u, t}$, and subsequently of $\Psi_{u}$ is only achieved when $t=u+1$ since at that stage all the possible expressions from which minimum is obtained are already examined and their minimum is stored in the auxiliary array of $M_{u,t}$ which gives the value of  $\varphi_{u, t}$.
At step 3.2.1 the operation on the lists   $\hat{Q}_{t-\nu(\mathbf{n}), t}^{\ell}(\mathbf{n})$  and  ${Q}_{t-\nu(\mathbf{n}), t}^{\ell}(\mathbf{n})$ for any two consecutive values of  $\mathbf{n_i},\mathbf{n_{i+1}}  \in \tilde\Pi_{t}'$ can be described as follows:
The first product arrangement in each of the lists for $1\leq\ell\leq m$ can be used to compute $\varphi_{t-\nu(\mathbf{n_i}), t}(\mathbf{n_i})$ and subsequently $M_{t-\nu(\mathbf{n_i}),t}$ using Ou's lemma since it is the vector in $V_{t}^{\ell}(\mathbf{n_i})$ with the lowest $G_{t, \ell}(\mathbf{N})$ value. Afterward, each of the lists needs to be updated which can be done firstly by applying the set maintenance procedure and then by applying to each of the elements in each list starting from the first element the check-and-remove procedure so as to retain only the optimal elements in each of the lists for $\mathbf{n_{i+1}}  \in \tilde\Pi_{t}'$.
Since each of the elements that are removed from a  $\hat{Q}_{t-\nu(\mathbf{n}), t}^{\ell}(\mathbf{n})$ can't reappear in it or any other list with $\ell'\geq\ell$ with the same being applied to the elements of lists  ${Q}_{t-\nu(\mathbf{n}), t}^{\ell}(\mathbf{n})$ the updating procedure takes a total of $O(m(R_1+R_2..R_{\vert\tilde{\Pi_t}'\vert})+\vert\hat{\Pi_t}'\vert)=O(R_1+R_2+..+R_{\vert\tilde{\Pi_t}\vert}+\vert\hat{\Pi_t}'\vert)=O(\vert\hat{\Pi_t}'\vert)=O(T^{m+1})$ steps for the check and removal of elements from each of the lists where $R_i$ is the number of removals of elements in the lists done for the $\mathbf{n_i}$ step since each element can be removed at most $m$ times due to the property of monotonic rebordering of the sets represented by the lists. It additionally takes   $O(m(R_1+R_2..R_{\vert\tilde{\Pi_t}'\vert})+\vert\hat{\Pi_t}'\vert)=O(R_1+R_2+..+R_{\vert\tilde{\Pi_t}\vert}+\vert\hat{\Pi_t}'\vert)=O(\vert\hat{\Pi_t}'\vert)=O(T^{m+1})$ steps for list insertions since an element can be inserted into a new list at most once and can never be reinserted at a list where it was removed before assuming that $t$ is fixed also due to the monotonic reordering property.
At Step 3.3 since we have both the elements $\mathbf{n}\in \tilde\Pi_{t}'$  and $\mathbf{N}\in \hat\Pi_{t}'$ for each $t$ in sorted order as well as all the necessary components to calculate $G_{t, \ell}(\mathbf{N})$ for each $t$ since all the relevant components have already been computed previously by our method we can put the elements $\mathbf{N}  \in \hat\Pi_{t}'$ for $\mathbf{n_1}  \in \tilde\Pi_{t}'$ in the relevant lists  $\hat{Q}_{t-\nu(\mathbf{n}), t}^{\ell}(\mathbf{n})$ and  ${Q}_{t-\nu(\mathbf{n}), t}^{\ell}(\mathbf{n})$ in increasing order in linear time and then apply the check-and-remove procedure the same way as previously for each element besides the first starting from the second in each of the lists ${Q}_{t-\nu(\mathbf{n}), t}^{\ell}(\mathbf{n})$ to have their elements in increasing $G_{t, \ell}(\mathbf{N})$ order. 
This step can be analyzed similarly to the previous use of the check-and-remove procedure since each element takes $O(R_i)$ time where $R_i$ is the number of elements removed from the list by it and given that $R_1+R_2+..R_q=O(T^{m+1})$   due to each element being removed at most once, the total complexity of this step is $O(T^{m+1})$.
\section{Appendix B}

For the case where there is no fractional period between the periods $u..v$ the following formula can be used 
$$\Psi_{u}=\min _{\substack{\mathbf{n} \in \Pi_{T-u+1} \\ w(\mathbf{n})=D(u, T)}}\left\{\bar{f}_{u, T}(\mathbf{n})\right\}$$ which requires $O\left(T^{m}\right)$ time.

For the case where $t=v$ we can use the following formula since by Lemma 1, we know that any period $i\in\{u, u+1, \ldots, t-1\}$ is a $B_{\tau}$-period for some $\tau \in\{0,1, \ldots, m\}$ if $t$ $>u$ :

\begin{equation*}
\begin{aligned}
\Psi_{u}=\min _{t=u, u+1, \ldots, T} \min _{\substack{\mathbf{n} \in \Pi_{t-u} \\ w(\mathbf{n})<D(u, t)}}\{\bar{f}_{u, t-1}(\mathbf{n})+P_{t}(D(u, t)-\\\omega(\mathbf{n}))+\Psi_{t+1}\}\end{aligned}\end{equation*}
The formula can be computed in $O\left(T^{m+1}\right)$ time.
Finally for the case where $t=u<v$ we can use the following formula:
since $t$ is a regeneration period and from the definition of $\Gamma_{u}(\mathbf{N})$ we have

$$\Psi_{u}=\min _{u<v \leq T} \min _{\substack{\mathbf{N} \in \Pi_{v-u-1} \\ w(\mathbf{N})<D(u, v)}}\left\{P_{u}(D(u, v)-\omega(\mathbf{N}))+F_{u}(\mathbf{N})\right\}$$
which has time complexity $O\left(T^{m+1}\right)$.
\bibliographystyle{cas-model2-names}

\bibliography{cas-refs}

\end{document}